\documentclass[graybox, envcountchap]{svmult}
\usepackage{mathptmx}        
\usepackage{helvet}          
\usepackage{courier}         
\usepackage{makeidx}         
\usepackage{graphicx}        
\usepackage{multicol}        
\usepackage[bottom]{footmisc}
\usepackage{rotating}
\usepackage{url}        
\usepackage{bm}
\usepackage{array}
\usepackage{units}
\usepackage{amsmath}
\usepackage{amssymb}
\usepackage{longtable}
\usepackage{dcolumn}
\begin{document}
\title*{Astrophysical Probes of Fundamental Physics}
\author{C.J.A.P. Martins}
\institute{C.J.A.P. Martins \at Centro de Astrof\'{\i}sica, Universidade do Porto, Rua das Estrelas, 4150-762 Porto, Portugal, \\ 
        and DAMTP/CTC, University of Cambridge, Wilberforce Road, Cambridge CB3 0WA, U.K., \email{Carlos.Martins@astro.up.pt}}
%
%
\maketitle

\abstract*{The dramatic confrontation between new observations and theories of the early and recent universe makes cosmology one of the most rapidly advancing fields in the physical sciences. The universe is a unique laboratory in which to probe fundamental physics, the rationale being to start from fundamental physics inspired models and explore their consequences in sufficient quantitative detail to be able to identify key astrophysical and cosmological tests of the underlying theory (or developing new tests when appropriate). An unprecedented number of such tests will be possible in the coming years, by exploiting the ever improving observational data. In this spirit I will highlight some open issues in cosmology and particle physics and provide some motivation for this symposium.}

\abstract{The dramatic confrontation between new observations and theories of the early and recent universe makes cosmology one of the most rapidly advancing fields in the physical sciences. The universe is a unique laboratory in which to probe fundamental physics, the rationale being to start from fundamental physics inspired models and explore their consequences in sufficient quantitative detail to be able to identify key astrophysical and cosmological tests of the underlying theory (or developing new tests when appropriate). An unprecedented number of such tests will be possible in the coming years, by exploiting the ever improving observational data. In this spirit I will highlight some open issues in cosmology and particle physics and provide some motivation for this symposium.}

\section{Scalar fields}

The deepest enigma of modern physics is whether or not there are fundamental scalar fields in nature. For over four decades the standard model of particle physics has been relying on one such field (the Higgs field) to give mass to all the other particles and make the theory gauge-invariant, but this hasn't yet been found. Finding it is the main science driver behind the LHC.

Despite the considerable success of the standard model of particle physics, there are at least three firmly established facts that it can't explain: neutrino masses, dark matter and the size of the baryon asymmetry of the universe. It's our confidence in the standard model that leads us to the expectation that there must be new physics beyond it. More importantly, all those three have obvious astrophysical and cosmological implications, so progress in fundamental particle physics will increasingly rely on progress in cosmology.

In this context, it is remarkable that Einstein gravity has no scalar fields. Indeed this is exceptional, because almost any consistent gravitational theory one can think of will have one or more scalar fields. The fact that there is none is, to some extent, what defines Einstein gravity. Nevertheless recent developments suggest that scalar fields can be equally important in astrophysics and cosmology.

Scalar fields play a key role in most paradigms of modern cosmology. One reason for their popularity is that they can take a VEV while preserving Lorentz invariance, while vector fields or fermions would break Lorentz Invariance and lead to conflicts with Special Relativity. Among other things, scalar fields have been invoked to account for the exponential expansion of the early universe (inflation), cosmological phase transitions and their relics (cosmic defects), dynamical dark energy powering current acceleration phase, and the variation of nature's fundamental couplings.

Even more important than each of these paradigms is the fact that they usually don't occur alone. For example, in most realistic inflation models the inflationary epoch ends with a phase transition at which defects are produced. Another example, to which we shall return later, is that dynamical scalar fields coupling to the rest of the model (which they will do, in any realistic scenario \cite{Carroll}) will necessarily have significant variations of fundamental constants. These links will be crucial for future consistency tests.

\section{Varying constants}

Nature is characterized by a set of physical laws and fundamental dimensionless couplings, which historically we have assumed to be spacetime-invariant. For the former this is a cornerstone of the scientific method, but for latter it is a simplifying assumption without further justification. Since it's these couplings that ultimately determine the properties of atoms, cells, planets and the universe as a whole, it's remarkable how little we know about them, and indeed that there is such a broad range of opinions on the subject \cite{Duff,Roysoc}.

At one side of the divide there is the Russian school which tends to see constants as defining asymptotic states (for example the speed of light is the limit velocity of massive particle in flat space-time, and so forth). At the opposite end of the spectrum is the Eddington school, which sees them as simple conversion factors. You are subscribing to the Eddington school when you 'set constants to unity' in your calculations. However, this can't be pushed arbitrarily far. One is free to choose units in which $c=\hbar=G=1$, but one can't choose units in which $c=\hbar=e=1$, because in the latter case the fine-structure constant will also be unity, and in the real world it isn't.

From the point of view of current physics, three constants seem fundamental: one only needs to define units of length, time and energy to carry out any experiment one chooses. However, nobody knows what will happen in a more fundamental theory? Will we still have three, one or none at all? (To some extent one can argue that in string theory there are only two fundamental constants). Moreover, it is also unknown whether they will be fixed by consistency conditions, or remain arbitrary (with their evolution being driven by some dynamical attractor mechanism). 

Although for model-building it makes sense to discuss variations of dimensional quantities, experimentally the situation is clear: one can only measure unambiguously dimensionless combinations of constants, and any such measurements are necessarily local. This is important if one wants to compare measurements at different cosmological epochs (searching for time variations) or even at different environments (searching for spatial variations). Face-value comparisons of measurements at different redshifts are too naive, and often manifestly incorrect. Most such comparisons are model-dependent, since a cosmological model must be assumed. In particular, assuming a constant rate of change (of the fine-structure constant say) is useless: no sensible particle physics model will ever have such dependence over any significant redshift range.

Thus speaking of variations of dimensional constants has no physical significance: one can concoct any variation by defining appropriate units of length, time and energy. Nevertheless, one is free to choose an arbitrary dimensionful unit as a standard and compare it with other quantities. A relevant example is the following. If one assumes particle masses to be constant, then constraints on the gravitational constant $G$ are in fact constraining the (dimensionless) product of $G$ and the nucleon mass squared. A better route is to compare the QCD interaction with the gravitational one: this can be done by assuming a fixed energy scale for QCD and allowing a varying $G$, or vice-versa. With these caveats, probes of a rolling $G$ provide key information on the gravitational sector. Paradoxically, $G$ was the first constant to be measured but is now the least well known, a consequence of the weakness of gravity.

If fundamental couplings are spacetime-varying, all the physics we know is incomplete and requires crucial revisions. Varying non-gravitational constants imply a violation of the Einstein Equivalence Principle, a fifth force of nature, and so on. Such a detection would therefore be revolutionary, but even improved null results are very important, and can provide constraints on anything from back-of-the-envelope toy models to string theory itself.

A simple way to illustrate the above point is as follows. If one imagines a cosmologically evolving scalar field that leads to varying fundamental couplings, then the natural timescale for their evolution would be the Hubble time. However, current local bounds from atomic clocks \cite{Rosenband} are already six orders of magnitude stronger. Any such field must therefore be evolving much more slowly that one would naively expect, and this rules out many otherwise viable models.

\section{Exciting times}

Searches for spacetime variations of fundamental constants mostly focus on the fine-structure constant $\alpha$ and the proton-to-electron mass ratio $\mu$, as the rest of these proceedings illustrate. These are the only two dimensionless parameters needed for the description of the gross structure of atoms and molecules. Additionally, the gravitational constant $G$ is interesting: although it has dimensions, one can construct dimensionless quantities involving it that probe the nature of the gravitational interaction (cf. Garcia-Berro's contribution in these proceedings).

It must be emphasized that in any sensible model where one of the couplings vary all the others should do as well, at some level, Different models predict very different relations for these variations; for example, in Grand-Unified Theories the variations of $\alpha$ and $\mu$ are related via
\begin{equation}
\frac{d\ln{\alpha}}{dt} = R \frac{d\ln{\alpha}}{dt}\,,
\end{equation}
where $R$ is a constant free parameter. Not even its size is determined a priori, although the naive expectation (based on high-energy GUT scenarios) is that the modulus of R should be of order 30-50. (Having R of order unity would require fine-tuning.) Hence, simultaneous astrophysical measurements of $\alpha$ and $\mu$, such as those in the radio band, provide an optimal way of probing GUTs and fundamental physics. Assuming the validity of the current claimed detections, measurements of $\alpha$ and $\mu$ with $10^{-6}$ accuracy can constrain R to $10\%$ (or better, depending on its value). 

Local (laboratory) bounds on the current rates of change can be obtained by looking at the frequency drift of two or more atomic clocks---see the contributions by Bize and Peik. These are currently consistent with no variation, and as has already been pointed out they note that these already restrict any variation to be many orders of magnitude weaker than the natural expectation for a cosmological process.

In coming years experiments such as $\mu$SCOPE, ACES, and possibly GG and STEP will carry out these measurements, as well as stringent Equivalence Principle tests, in microgravity conditions. These will improve on current sensitivities by several orders of magnitude, and if the current claims of astrophysical detections are correct they should find direct evidence for Equivalence Principle violations.

A nominally strong bound on $\alpha$ can be obtained from the Oklo natural nuclear reactor. However, this is obtained assuming that $\alpha$ is the only quantity that can change, which is a particularly poor assumption since it is known that the underlying chain or reactions is most sensitive to the coupling for the strong force (cf. Flambaum's contribution). The importance of this Oklo bound has therefore been grossly exaggerated.

Astrophysical measurements rely on precision spectroscopy, most often using absorption lines. Emission lines can also be used in principle, but with current means they are less sensitive---cf. the contribution by Gutierrez. Note however that although it's often claimed that another disadvantage of emission line measurements is that they can only be done at low redshift, this has been shown not to be the case \cite{Brinchmann}.

For both $\alpha$ and $\mu$ there are few-sigma claims of detected variations \cite{Murphy,Ubachs} at redshfits $z\sim3$, but in both cases other studies find no variations. Recent months have seen the emergence of interesting new results, including evidence for a spatial dipole and further spatial variations within the Galaxy, and this symposium provided a timely opportunity for those involved to present and discuss them---cf. the contributions by Levshakov, Petitjean, Thompson, Webb, Wendt and Ubachs.

Bounds on $\alpha$ can be obtained at much higher redshift, in particular using the Cosmic Microwave Background \cite{Martins}. WMAP data, when combined with other datasets, has recently led to constrains on any such variation to be below the percent level---cf. Galli's contribution. There is no evidence for variation at this redshift. The bound is much weaker than the above (due to degeneracies with other cosmological parameters) but clean and model-independent. At even higher redshifts BBN can also be used (and also yields percent-level bounds), although with the caveat that in these case the bounds are necessarily model-dependent. The CMB data is also becoming good enough to constrain joint variations of $\alpha$ and other couplings. Such joint constraints will become increasingly important.

\section{Redundancy and further tests}

This field is perceived by outsiders as one plagued with controversy and haunted by concerns about systematics. Although this view is certainly unfair, it is true that improvements are needed---but it most also be emphasized that those active in the field are the first to admit this. Part of the problem stems from the fact that almost all the data that has been used so far was gathered for other purposes, and does not have the necessary quality to fully exploit the capabilities of modern spectrographs. Measurement of fundamental constants requires observing procedures beyond what is done in standard observations \cite{Thompson}. One needs customized data acquisition and wavelength calibration procedures beyond those supplied by standard pipelines, and ultimately one should calibrate with laser frequency combs. Fortunately, we are moving fast in this direction.

A new generation of high-resolution, ultra-stable spectrographs will be needed to resolve the issue. In the short term Maestro at the MMT and PEPSI at LBT will being significant improvements. Later on the prospects are even better with two ESO spectrographs, ESPRESSO for VLT (which has recently been approved by the ESO Council) and CODEX for the E-ELT. Both of these have searches for these variations as one of the key science drivers, and will improve on currently achievable sensitivities by one and two orders of magnitude, respectively.(Further details can be found in Molaro's contribution.) In the meantime, and ongoing VLT/UVES Large Programme will bring significant improvements.

With the anticipated gains if both statistical and systematic uncertainties, further tests will also become possible. For example, the ratio of the proton and electron masses $\mu$ is measured through molecular vibrational and rotational lines. If one uses $H_2$ then one is indeed measuring $\mu$, but molecular Hydrogen is not easy to find, and therefore other molecules are also used, and in this case one needs to be more careful. Strictly speaking one is then probing an average nucleon mass containing both protons and neutrons. One can still write $m_{nuc}/m_e\sim F m_p/m_e$ (with $F$ being a number of order a few), but only if there are no composition-dependent forces, or in other words if the scalar field has the same coupling to protons and neutrons. However, from a theoretical point of view this is highly unlikely. Nevertheless, this provides us with a golden opportunity to search for these couplings. All that needs to be done is to repeat the measurements with molecules containing different numbers of neutrons: $H_2$ has none, $HD$ has one, and so forth. It should also be added that one can sometimes find these various molecules in the same system, so this need not be costly in terms of telescope time.

Moreover, given the extraordinary relevance of possible detections, it is especially important to have these confirmed by alternative and completely independent methods. Over the past decade or so a whole range of experimental and observational techniques have been used to search for temporal and spatial varying couplings throughout the cosmic history. Atomic clocks, geophysics (Oklo and meteorites), spectroscopy, the CMB and BBN have already been mentioned---they are the best known and more often used ones. But further astrophysical probes are also emerging, and they can play a key role in the coming years: examples are clusters (through the Sunyaev-Zel'dovich effect), helioseismology and strong gravity systems (white dwarfs and neutron stars).

Last but not least, complementary tests will be crucial to establish the robustness and consistency of the results. These are tests which do not measure varying constants directly but search for other non-standard effects that must be present if constants do vary. Equivalence Principle tests are the best known example. A second one that may well be crucial in the coming years are tests of the temperature-redshift relation. For example, in many models where photons are destroyed one car write
\begin{equation}
T(z)=T_0(1+z)^{1-\beta}\,,
\end{equation}
where $\beta=0$ in the standard model, and the current best constraint is $\beta<0.08$ \cite{Luzzi}.

Measuring the CMB temperature at non-zero redshift is not trivial, but the systems where it can be done are also interesting for varying constants. At low redshifts $T(z)$ can be measured at SZ clusters \cite{Battistelli}, which can also be used to measure $\alpha$, and at intermediate redshifts it can be measured spectroscopically using molecular rotational transitions \cite{Srianand}, which can also be used to measure $\mu$. The prospect of simultaneous measurements of $T(z)$, $\mu$ and possibly also $\alpha$ in the same system, with ESPRESSO or CODEX, is a particularly exciting one.

\section{Dynamical dark energy}

Observations suggest that the universe dominated by component whose gravitational behavior is similar to that of a cosmological constant. The required cosmological constant value is so small that a dynamical scalar field is arguably more likely. The fact that it must be slow-rolling in recent times (which is mandatory for $p<0$) and dominating the energy budget are sufficient to ensure \cite{Carroll} that couplings of this field lead to observable long-range forces and time dependence of the constants of nature. It should be kept in mind that in any sensible theory scalars will couple to the rest of the world in any manner not prevented by symmetry principles.

Standard methods (SNe, Lensing, etc) are know to be of limited use as dark energy probes \cite{Maor}. One reason for this is that what one observes and what one wants to measure are related by second derivatives. A clear detection of varying dark energy equation of state $w(z)$ is key to a convincing result. Since $w_0\sim -1$ and since the field is slow-rolling when dynamically important, a convincing detection of $w(z)$ is quite unlikely even with EUCLID or WFIRST.

Since a scalar field yielding dark energy also yields varying couplings, they can be used to reconstruct $w(z)$ \cite{Lidsey}. The procedure is analogous to reconstructing the 1D potential for the classical motion of a particle, given its trajectory. The simplest paradigm relating the two is
\begin{equation}
\frac{\Delta\alpha}{\alpha}=\kappa\zeta(\delta\phi)\,,
\end{equation}
where $\kappa^2=8\pi G$. This reconstruction method only involves first derivatives of the data, and it will complement and extend traditional methods. A comparison of this and the standard method will yield a measurement of the scalar field coupling $\zeta$, which can be compared to that coming from Equivalence Principle tests. In the E-ELT era, synergies will also exist with the Sandage-Loeb test \cite{Sandage}.

Advantages of this method include the fact that it allows direct probes of Grand Unification and fundamental physics, and that it directly distinguishes a cosmological constant from a dynamical field (with no false positives). However, the key advantage is its huge redshift lever arm, probing the otherwise inaccessible redshift range where the field dynamics is expected to be fastest (that is, deep in the matter era). It is of course also much cheaper than putting a satellite in space: it is a ground-based method, and taking at face value the currently existing data one can show \cite{Pipeline} that 100 good nights on a 10m-class telescope (such as the VLT, Keck or the LBT) could conceivably yield a five-sigma detection of dynamical dark energy.

\section{Conclusions}

Varying constants are a powerful, versatile and low-cost way to probe fundamental physics and dark energy. There is ample experimental evidence showing that fundamental couplings run with energy, and many particle physics and cosmology models suggest that they also roll with time. There is therefore every incentive to search for these, and there's no better place than in the early universe. Current measurements restrict any such relative variations to be below the $10^{-5}$ level, which is already a very significant constraint.

The coming years will bring big gains in sensitivity and also dedicated experiments, but doing things right is tough: we need customized observation procedures, laser frequency comb calibration, purpose-built data reduction pipelines, and further astrophysical probes to complement the existing ones. One must also keep in mind the dark energy lesson: when measurements from type Ia supernovae first suggested an accelerating universe these were largely dismissed until such evidence also emerged through independent methods (CMB, lensing, large-scale structure and so on). It is this quest for redundancy that this field must now pursue, and laboratory measurements (with atomic clocks), Equivalence Principle and temperature-redshift tests will be crucial in the next decade.

In addition to its direct impact, these studies have a unique role to play in shedding light on the enigma of dark energy. The early universe is an ideal fundamental physics laboratory, allowing us to carry out tests that one will never be able to do in terrestrial laboratories. Recent technological developments now provide us with tools to accurately search for varying constants and explore its impacts elsewhere, and this opportunity must be taken. The fact that something as fundamental (and abstract) as string theory may one day be confirmed using something as mundane as spectroscopy is an opportunity that neither astrophysicists nor particle physicists can afford to miss.

\begin{acknowledgement}
I acknowledge discussions with a number of collaborators and colleagues---Silvia Galli, Gemma Luzzi, Jo\~ao Magueijo, Elisabetta Majerotto, Gianpiero Mangano, Alessandro Melchiorri, Eloisa Menegoni, Paolo Molaro, Michael Murphy, Nelson Nunes, Pedro Pedrosa, Patrick Petitjean, Georg Robbers, Rodger Thompson and Wim Ubachs---which have shaped my views on this subject. This work is funded by a Ci\^encia2007 Research Contract, supported by FSE and POPH-QREN funds. Further support came from Funda\c c\~ao para a Ci\^encia e a Tecnologia (FCT), Portugal, through grant PTDC/CTE-AST/098604/2008.
\end{acknowledgement}


\begin{thebibliography}{99.}%
\bibitem{Pipeline} P.P. Avelino, C.J.A.P. Martins, N.J. Nunes and K.A. Olive, Phys. Rev. D74 (2006) 083508.
\bibitem{Battistelli} E.S. Battistelli et al., Astrophys. J. 580 (2002) L101. 
\bibitem{Brinchmann} J. Brinchmann et al., AIP Conf. Proc. 736 (2005) 117.
\bibitem{Carroll} S.M. Carroll, Phys. Rev. Lett. 81 (1998) 3067.
\bibitem{Duff} M.J. Duff, L.B. Okun and G. Veneziano, JHEP 03 (2002) 023.
\bibitem{Maor} I. Maor, R. Brustein, J. McMahon and P.J. Steinhardt, Phys. Rev. D65 (2002) 123003.
\bibitem{Roysoc} C.J.A.P. Martins, Phil. Trans. Roy. Soc. Lond. A360 (2002) 2681.
\bibitem{Martins} C.J.A.P. Martins et al., Phys. Lett. B585 (2004) 29.
\bibitem{Murphy} M.T. Murphy, J.K. Webb and V.V.Flambaum, Mon. Not. Roy. Astron. Soc. 345 (2003) 609.
\bibitem{Lidsey} N.J. Nunes and J.E. Lidsey, Phys. Rev. D69 (2004) 123511.
\bibitem{Luzzi} G. Luzzi et al., Astrophys. J. 705 (2009) 1122.
\bibitem{Ubachs} E. Reinhold et al., Phys. Rev. Lett. 96 (2006) 151101.
\bibitem{Rosenband} T. Rosenband et al., Science 319 (2008) 1808. 
\bibitem{Sandage} A. Sandage, Astrophysical Journal 136 (1962) 319.
\bibitem{Srianand} R. Srianand, P. Petitjean and C. Ledoux, Nature 408 (2000) 931.
\bibitem{Thompson} R.I. Thompson, J. Bechtold, J.H. Black and C.J.A.P. Martins, New Astron. 14 (2009) 379.
\end{thebibliography}
\end{document}